\documentclass[10pt,fleqn]{article}
\usepackage{amsfonts,amsmath,amssymb,cite}
\newcommand{\sgn}{\textrm{sgn}}

\begin{document}
\title{Recurrence and differential relations for spherical spinors}
\author{Rados{\l}aw Szmytkowski \\*[3ex]
Atomic Physics Division, 
Department of Atomic Physics and Luminescence, \\
Faculty of Applied Physics and Mathematics, 
Gda{\'n}sk University of Technology, \\
Narutowicza 11/12, PL 80--233 Gda{\'n}sk, Poland \\
email: radek@mif.pg.gda.pl}
\date{}
\maketitle
\begin{center}
\textbf{Published as: J.\ Math.\ Chem.\ 42 (2007) 397--413} \\*[1ex]
\textbf{doi: 10.1007/s10910-006-9110-0} \\*[5ex]
\end{center}
\begin{abstract} 
We present a comprehensive table of recurrence and differential
relations obeyed by spin one-half spherical spinors (spinor spherical
harmonics) $\Omega_{\kappa\mu}(\mathbf{n})$ used in relativistic
atomic, molecular, and solid state physics, as well as in
relativistic quantum chemistry. First, we list finite expansions in
the spherical spinor basis of the expressions
$\mathbf{A}\cdot\mathbf{B}\,\Omega_{\kappa\mu}(\mathbf{n})$ and
\mbox{$\mathbf{A}\cdot(\mathbf{B}\times\mathbf{C})\,
\Omega_{\kappa\mu}(\mathbf{n})$}, where $\mathbf{A}$, $\mathbf{B}$,
and $\mathbf{C}$ are either of the following vectors or vector
operators: $\mathbf{n}=\mathbf{r}/r$ (the radial unit vector),
$\mathbf{e}_{0}$, $\mathbf{e}_{\pm1}$ (the spherical, or cyclic,
versors), $\boldsymbol{\sigma}$ (the $2\times2$ Pauli matrix vector),
$\hat{\mathbf{L}}=-i\mathbf{r}\times\boldsymbol{\nabla}I$ (the
dimensionless orbital angular momentum operator; $I$ is the
$2\times2$ unit matrix),
$\hat{\mathbf{J}}=\hat{\mathbf{L}}+\frac{1}{2}\boldsymbol{\sigma}$
(the dimensionless total angular momentum operator). Then, we list
finite expansions in the spherical spinor basis of the expressions
$\mathbf{A}\cdot\mathbf{B}\,F(r)\Omega_{\kappa\mu}(\mathbf{n})$ and
$\mathbf{A}\cdot(\mathbf{B}\times\mathbf{C})\,
F(r)\Omega_{\kappa\mu}(\mathbf{n})$, where at least one of the
objects $\mathbf{A}$, $\mathbf{B}$, $\mathbf{C}$ is the nabla
operator $\boldsymbol{\nabla}$, while the remaining ones are chosen
from the set $\mathbf{n}$, $\mathbf{e}_{0}$, $\mathbf{e}_{\pm1}$,
$\boldsymbol{\sigma}$, $\hat{\mathbf{L}}$, $\hat{\mathbf{J}}$.


\end{abstract}
%
%
\section{Introduction}
\label{I}
\setcounter{equation}{0}
The spin one-half spherical spinors (spinor spherical harmonics)
emerge in relativistic quantum mechanics in the context of the
separation of spherical variables when solving the central-field
Dirac problem, e.g., in the relativistic analysis of one-electron
atoms \cite{Akhi59,Rose61,Bjor64,Grei90,Thal92}. In the relativistic
theory of many-electron systems (including atoms, molecules, and the
solid state), they enter elementary one-electron Dirac central-field
orbitals of which approximate multi-electron wave functions
describing such systems are frequently constructed (see numerous
references cited in Refs.\ \cite{Pyyk86,Pyyk93,Pyyk00,RTAM}).

Despite the so well documented practical importance of the spherical
spinors, thus far relatively little space has been devoted in the
literature to systematic studies or presentations of their
properties. Standard textbooks or reference works on the angular
momentum theory, such as Refs.\ \cite{Rose57,Edmo60,Bied81,Louc96},
discuss the spherical spinors only superficially. In fact, even in
the most comprehensive relevant treatise by Varshalovich \emph{et
al.\/} \cite{Vars75} the spherical spinors have been treated much
less exhaustively than scalar or vector spherical harmonics. In
consequence, atomic and molecular researchers usually have to derive
particular properties of the spherical spinors \emph{ad hoc}, as
exemplified by Refs.\ \cite{Bech93,Rutk97,Szmy02,Szmy04,Miel06}. As a
part of our research program (in this connection, see also Ref.\
\cite{Szmy05}) aimed at changing this unsatisfactory situation, in
this paper we present a comprehensive table of recurrence and
differential relations obeyed by the spherical spinors.
%
%
\section{Preliminaries}
\label{II}
\setcounter{equation}{0}
\subsection{Definitions}
\label{II.1}
Let $\mathbf{e}_{x}$, $\mathbf{e}_{y}$, $\mathbf{e}_{z}$ be versors
of some right-handed Cartesian coordinate system. The cyclic versors
$\mathbf{e}_{0}$, $\mathbf{e}_{\pm1}$, are defined through the
relationships
\begin{equation}
\mathbf{e}_{0}=\mathbf{e}_{z},
\qquad 
\mathbf{e}_{\pm1}
=\mp\frac{1}{\sqrt{2}}(\mathbf{e}_{x}\pm i\mathbf{e}_{y}).
\label{2.1.1}
\end{equation}
The dimensionless orbital ($\hat{\mathbf{L}}$) and total
($\hat{\mathbf{J}}$) angular momentum operators (both with respect to
the center of the aforementioned Cartesian system) are defined as
\begin{equation}
\hat{\mathbf{L}}=-i\mathbf{r}\times\boldsymbol{\nabla}I
\label{2.1.2}
\end{equation}
and
\begin{equation}
\hat{\mathbf{J}}=\hat{\mathbf{L}}
+{\textstyle\frac{1}{2}}\boldsymbol{\sigma},
\label{2.1.3}
\end{equation}
respectively. In Eqs.\ (\ref{2.1.2}) and (\ref{2.1.3}), and
hereafter, $I$ is the $2\times2$ unit matrix, while
$\boldsymbol{\sigma}$ is the $2\times2$ Pauli matrix vector
\begin{equation}
\boldsymbol{\sigma}=\sigma_{x}\mathbf{e}_{x}+\sigma_{y}\mathbf{e}_{y}
+\sigma_{z}\mathbf{e}_{z},
\label{2.1.4}
\end{equation}
with
\begin{equation}
\sigma_{x}
=\left(
\begin{array}{cc}
0 & 1 \\
1 & 0
\end{array}
\right),
\qquad
\sigma_{y}
=\left(
\begin{array}{cc}
0 & -i \\
i & 0
\end{array}
\right),
\qquad
\sigma_{z}
=\left(
\begin{array}{cc}
1 & 0 \\
0 & -1
\end{array}
\right).
\label{2.1.5}
\end{equation}

Let $\mathbf{n}=\mathbf{r}/r$ be a unit radius vector with respect to
the origin of the aforementioned Cartesian system. The spatial
orientation of $\mathbf{n}$ is uniquely determined by specifying two
angles, $0\leqslant\theta\leqslant\pi$ and $0\leqslant\varphi<2\pi$,
such that
\begin{equation}
\mathbf{e}_{x}\cdot\mathbf{n}=\sin\theta\cos\varphi,
\qquad
\mathbf{e}_{y}\cdot\mathbf{n}=\sin\theta\sin\varphi,
\qquad
\mathbf{e}_{z}\cdot\mathbf{n}=\cos\theta.
\label{2.1.6}
\end{equation}
Evidently, $\theta$ and $\varphi$ are, respectively, the polar and
the azimuthal angles in the spherical system of coordinates, with its
polar and azimuthal axes directed along the Cartesian versors
$\mathbf{e}_{z}$ and $\mathbf{e}_{x}$, respectively.

We define the spin one-half spherical spinors, hereafter denoted as
$\Omega_{\kappa\mu}(\mathbf{n})$, as two-component functions of the
unit vector $\mathbf{n}$ (or, equivalently, of the aforementioned
angles $\theta$ and $\varphi$) of the form
\begin{equation}
\Omega_{\kappa\mu}(\mathbf{n})
=\left(
\begin{array}{c}
\sgn(-\kappa)\sqrt{\frac{\kappa+\frac{1}{2}-\mu}{2\kappa+1}}\:
Y_{l,\mu-1/2}(\mathbf{n}) \\*[1ex]
\sqrt{\frac{\kappa+\frac{1}{2}+\mu}{2\kappa+1}}\:
Y_{l,\mu+1/2}(\mathbf{n})
\end{array}
\right),
\label{2.1.7}
\end{equation}
with $\kappa\in\{\pm1,\pm2,\ldots\}$,
$\mu\in\{-|\kappa|+\frac{1}{2},-|\kappa|+\frac{3}{2},\ldots,
|\kappa|-\frac{1}{2}\}$, and
\begin{equation}
l=|\kappa+{\textstyle\frac{1}{2}}|-{\textstyle\frac{1}{2}}
=\left\{
\begin{array}{ll}
\kappa & \textrm{for $\kappa>0$} \\
-\kappa-1 & \textrm{for $\kappa<0$}.
\end{array}
\right.
\label{2.1.8}
\end{equation}
In Eq.\ (\ref{2.1.7}),
\begin{equation}
Y_{lm}(\mathbf{n})=\sqrt{\frac{2l+1}{4\pi}\frac{(l-m)!}{(l+m)!}}\:
P_{l}^{(m)}(\cos\theta)\mathrm{e}^{im\varphi}
\label{2.1.9}
\end{equation}
is the scalar spherical harmonics, with
\begin{equation}
P_{l}^{(m)}(\xi)=\frac{(-)^{m}}{2^{l}l!}(1-\xi^{2})^{m/2}
\frac{\mathrm{d}^{l+m}}{\mathrm{d}\xi^{l+m}}(\xi^{2}-1)^{l}
\qquad (-1\leqslant\xi\leqslant+1)
\label{2.1.10}
\end{equation}
being the associated Legendre function of the first kind. (The phases
in Eqs.\ (\ref{2.1.9}) and (\ref{2.1.10}) have been chosen so that
the spherical harmonics (\ref{2.1.9}) conforms to the
Condon--Shortley \cite{Cond35} phase convention; in this connection,
see also the remark concluding Sec.\ \ref{II.2}.)

In this work, we shall label the spherical spinors with the two
indices $\kappa$ and $\mu$. However, it should be mentioned that in
the relevant literature one encounters also numerous examples of
labeling these functions by three indices $j$, $l$, and $\mu$, with
the first index in this triple related to $\kappa$ through
\begin{equation}
j=|\kappa|-{\textstyle\frac{1}{2}},
\label{2.1.11}
\end{equation}
with $l$ defined as in Eq.\ (\ref{2.1.8}), and with $\mu$ assuming
the same value as explained below Eq.\ (\ref{2.1.7}). In view of the
relations (\ref{2.1.8}) and (\ref{2.1.11}), and the converse one,
\begin{equation}
\kappa=(l-j)(2j+1),
\label{2.1.12}
\end{equation}
both labeling schemes are completely equivalent.
\subsection{Remarks}
\label{II.2}
Preparing the collection of formulas presented in Sec.\ \ref{III} we
have made an attempt to minimize the number of entries (with a few
exceptions when the operator $\hat{\mathbf{J}}$ has been involved).
To this end, we have made an extensive use of the property
\begin{equation}
\mathbf{A}\cdot(\mathbf{B}\times\mathbf{C})
=(\mathbf{A}\times\mathbf{B})\cdot\mathbf{C},
\label{2.2.1}
\end{equation}
valid alike for ordinary vectors and vector operators. Also, we have
extensively exploited the identities like
\begin{equation}
\hat{\mathbf{L}}\times\hat{\mathbf{L}}=i\hat{\mathbf{L}},
\qquad
\boldsymbol{\sigma}\times\boldsymbol{\sigma}=2i\boldsymbol{\sigma},
\qquad
\hat{\mathbf{J}}\times\hat{\mathbf{J}}=i\hat{\mathbf{J}},
\label{2.2.2}
\end{equation}
\begin{equation}
\mathbf{n}\times\boldsymbol{\nabla}
=-\boldsymbol{\nabla}\times\mathbf{n},
\qquad
\mathbf{n}\cdot\hat{\mathbf{L}}=\hat{\mathbf{L}}\cdot\mathbf{n}=0,
\qquad
\boldsymbol{\nabla}\cdot\hat{\mathbf{L}}
=\hat{\mathbf{L}}\cdot\boldsymbol{\nabla}=0,
\label{2.2.3}
\end{equation}
etc., to reduce operators acting on the spherical spinors to the
simplest possible forms. If a result of such a reduction of a
particular operator has been found to be a scalar multiple of the
identity, the action of this operator on the spherical spinors has
not been displayed in the table. For instance, the equation
\begin{equation}
(\mathbf{n}\times\hat{\mathbf{L}})\times\mathbf{n}\,
\Omega_{\kappa\mu}(\mathbf{n})=2i\Omega_{\kappa\mu}(\mathbf{n})
\label{2.2.4} 
\end{equation}
has not been included in Sec.\ \ref{III} since it reflects the
operator identity
\begin{equation}
\mathbf{n}\times(\hat{\mathbf{L}}\times\mathbf{n})
=(\mathbf{n}\times\hat{\mathbf{L}})\times\mathbf{n}=2iI
\label{2.2.5}
\end{equation}
rather than some particular property of the spherical spinors.

In Sec.\ \ref{III.3}, $F(r)$ is a (once or twice, depending on the
needs) differentiable, and otherwise arbitrary, function of the
radial variable $r=|\mathbf{r}|$.

Before proceeding to the table, a word of caution is still in order.
It appears that, analogously to the case of scalar spherical
harmonics, recurrence and differential relations obeyed by the
spherical spinors depend on the choices of phases in the defining
equations. In other words, if the net phase of spherical spinors in
use differs from that following from our Eqs.\ (\ref{2.1.7}),
(\ref{2.1.9}), and (\ref{2.1.10}), or if the Pauli matrix
$\sigma_{y}$ is defined with the opposite sign, as it occasionally
happens in the literature, signs in some of the relationships listed
in Sec.\ \ref{III} may need to be changed.
%
%
\section{Table of recurrence and differential relations for spherical
spinors} 
\label{III}
\subsection{Algebraic recurrence relations}
\label{III.1}
\setcounter{equation}{0}
\begin{eqnarray}
\mathbf{e}_{0}\cdot\mathbf{n}\,
\Omega_{\kappa\mu}(\mathbf{n})
&=& -\frac{2\mu}{4\kappa^{2}-1}\Omega_{-\kappa\mu}(\mathbf{n})
+\frac{\sqrt{(\kappa+\frac{1}{2})^{2}-\mu^{2}}}{|2\kappa+1|}
\Omega_{\kappa+1,\mu}(\mathbf{n})
\nonumber \\
&& +\frac{\sqrt{(\kappa-\frac{1}{2})^{2}-\mu^{2}}}{|2\kappa-1|}
\Omega_{\kappa-1,\mu}(\mathbf{n})
\label{3.1.1}
\end{eqnarray}
\begin{eqnarray}
\mathbf{e}_{\pm1}\cdot\mathbf{n}\,
\Omega_{\kappa\mu}(\mathbf{n})
&=& \pm\sqrt{2}\frac{\sqrt{\kappa^{2}-(\mu\pm\frac{1}{2})^{2}}}
{4\kappa^{2}-1}\Omega_{-\kappa,\mu\pm1}(\mathbf{n})
\nonumber \\
&& +\frac{\sqrt{(\kappa\pm\mu+\frac{1}{2})
(\kappa\pm\mu+\frac{3}{2})}}{\sqrt{2}(2\kappa+1)}
\Omega_{\kappa+1,\mu\pm1}(\mathbf{n})
\nonumber \\
&& -\frac{\sqrt{(\kappa\mp\mu-\frac{1}{2})
(\kappa\mp\mu-\frac{3}{2})}}{\sqrt{2}(2\kappa-1)}
\Omega_{\kappa-1,\mu\pm1}(\mathbf{n})
\label{3.1.2}
\end{eqnarray}
\begin{equation}
\mathbf{n}\cdot\boldsymbol{\sigma}\,\Omega_{\kappa\mu}(\mathbf{n})
=-\Omega_{-\kappa\mu}(\mathbf{n})
\label{3.1.3}
\end{equation}
\begin{equation}
\mathbf{e}_{0}\cdot\boldsymbol{\sigma}\,\Omega_{\kappa\mu}(\mathbf{n})
=-\frac{2\mu}{2\kappa+1}\Omega_{\kappa\mu}(\mathbf{n})
-2\frac{\sqrt{(\kappa+\frac{1}{2})^{2}-\mu^{2}}}{|2\kappa+1|}
\Omega_{-\kappa-1,\mu}(\mathbf{n})
\label{3.1.4}
\end{equation}
\begin{eqnarray}
\mathbf{e}_{\pm1}\cdot\boldsymbol{\sigma}\,
\Omega_{\kappa\mu}(\mathbf{n})
&=& \pm\sqrt{2}
\frac{\sqrt{\kappa^{2}-(\mu\pm\frac{1}{2})^{2}}}{2\kappa+1}
\Omega_{\kappa,\mu\pm1}(\mathbf{n})
\nonumber \\
&& -\sqrt{2}\frac{\sqrt{(\kappa\pm\mu+\frac{1}{2})
(\kappa\pm\mu+\frac{3}{2})}}{2\kappa+1}
\Omega_{-\kappa-1,\mu\pm1}(\mathbf{n})
\label{3.1.5}
\end{eqnarray}
\begin{eqnarray}
\mathbf{e}_{0}\cdot(\mathbf{n}\times\boldsymbol{\sigma})\,
\Omega_{\kappa\mu}(\mathbf{n})
&=& i\frac{4\mu\kappa}{4\kappa^{2}-1}
\Omega_{-\kappa\mu}(\mathbf{n})
+i\frac{\sqrt{(\kappa+\frac{1}{2})^{2}-\mu^{2}}}{|2\kappa+1|}
\Omega_{\kappa+1,\mu}(\mathbf{n})
\nonumber \\
&& -i\frac{\sqrt{(\kappa-\frac{1}{2})^{2}-\mu^{2}}}{|2\kappa-1|}
\Omega_{\kappa-1,\mu}(\mathbf{n})
\label{3.1.6}
\end{eqnarray}
\begin{eqnarray}
\mathbf{e}_{\pm1}\cdot(\mathbf{n}\times\boldsymbol{\sigma})\,
\Omega_{\kappa\mu}(\mathbf{n})
&=& \mp i2\sqrt{2}\kappa
\frac{\sqrt{\kappa^{2}-(\mu\pm\frac{1}{2})^{2}}}{4\kappa^{2}-1}
\Omega_{-\kappa,\mu\pm1}(\mathbf{n})
\nonumber \\
&& +i\frac{\sqrt{(\kappa\pm\mu+\frac{1}{2})
(\kappa\pm\mu+\frac{3}{2})}}{\sqrt{2}(2\kappa+1)}
\Omega_{\kappa+1,\mu\pm1}(\mathbf{n})
\nonumber \\
&& +i\frac{\sqrt{(\kappa\mp\mu-\frac{1}{2})
(\kappa\mp\mu-\frac{3}{2})}}{\sqrt{2}(2\kappa-1)}
\Omega_{\kappa-1,\mu\pm1}(\mathbf{n})
\label{3.1.7}
\end{eqnarray}
%
%
\subsection{Differential relations of the first kind}
\label{III.2}
\setcounter{equation}{0}
\begin{equation}
\mathbf{e}_{0}\cdot\hat{\mathbf{L}}\,
\Omega_{\kappa\mu}(\mathbf{n})
=\frac{2\mu(\kappa+1)}{2\kappa+1}
\Omega_{\kappa\mu}(\mathbf{n})
+\frac{\sqrt{(\kappa+\frac{1}{2})^{2}-\mu^{2}}}{|2\kappa+1|}
\Omega_{-\kappa-1,\mu}(\mathbf{n})
\label{3.2.1}
\end{equation}
\begin{eqnarray}
\mathbf{e}_{\pm1}\cdot\hat{\mathbf{L}}\,\Omega_{\kappa\mu}(\mathbf{n})
&=& \mp\sqrt{2}(\kappa+1)
\frac{\sqrt{\kappa^{2}-(\mu\pm\frac{1}{2})^{2}}}
{2\kappa+1}\Omega_{\kappa,\mu\pm1}(\mathbf{n})
\nonumber \\
&& +\frac{\sqrt{(\kappa\pm\mu+\frac{1}{2})
(\kappa\pm\mu+\frac{3}{2})}}
{\sqrt{2}(2\kappa+1)}\Omega_{-\kappa-1,\mu\pm1}(\mathbf{n})
\label{3.2.2}
\end{eqnarray}
\begin{equation}
\boldsymbol{\sigma}\cdot\hat{\mathbf{L}}\,
\Omega_{\kappa\mu}(\mathbf{n})
=-(\kappa+1)\Omega_{\kappa\mu}(\mathbf{n})
\label{3.2.3}
\end{equation}
\begin{equation}
\mathbf{n}\cdot\hat{\mathbf{J}}\,\Omega_{\kappa\mu}(\mathbf{n})
=-\frac{1}{2}\Omega_{-\kappa\mu}(\mathbf{n})
\label{3.2.4}
\end{equation}
\begin{equation}
\mathbf{e}_{0}\cdot\hat{\mathbf{J}}\,\Omega_{\kappa\mu}(\mathbf{n})
=\mu\Omega_{\kappa\mu}(\mathbf{n})
\label{3.2.5}
\end{equation}
\begin{equation}
\mathbf{e}_{\pm1}\cdot\hat{\mathbf{J}}\,
\Omega_{\kappa\mu}(\mathbf{n})=\mp\frac{1}{\sqrt{2}}
\sqrt{\kappa^{2}-(\mu\pm{\textstyle\frac{1}{2}})^{2}}\,
\Omega_{\kappa,\mu\pm1}(\mathbf{n})
\label{3.2.6}
\end{equation}
\begin{equation}
\boldsymbol{\sigma}\cdot\hat{\mathbf{J}}\,
\Omega_{\kappa\mu}(\mathbf{n})
=-\left(\kappa-{\textstyle\frac{1}{2}}\right)
\Omega_{\kappa\mu}(\mathbf{n})
\label{3.2.7}
\end{equation}
\begin{equation}
\hat{\mathbf{L}}^{2}\,\Omega_{\kappa\mu}(\mathbf{n})
=\kappa(\kappa+1)\Omega_{\kappa\mu}(\mathbf{n})
\label{3.2.8}
\end{equation}
\begin{equation}
\hat{\mathbf{J}}^{2}\,\Omega_{\kappa\mu}(\mathbf{n})
=\left(\kappa^{2}-{\textstyle\frac{1}{4}}\right)
\Omega_{\kappa\mu}(\mathbf{n})
\label{3.2.9}
\end{equation}
\begin{equation}
\hat{\mathbf{L}}\cdot\hat{\mathbf{J}}\,\Omega_{\kappa\mu}(\mathbf{n})
=\left(\kappa-{\textstyle\frac{1}{2}}\right)(\kappa+1)
\Omega_{\kappa\mu}(\mathbf{n})
\label{3.2.10}
\end{equation}
\begin{eqnarray}
\mathbf{e}_{0}\cdot(\mathbf{n}\times\hat{\mathbf{L}})\,
\Omega_{\kappa\mu}(\mathbf{n})
&=& -i\frac{2\mu(\kappa+1)}{4\kappa^{2}-1}
\Omega_{-\kappa\mu}(\mathbf{n})
-i\kappa\frac{\sqrt{(\kappa+\frac{1}{2})^{2}-\mu^{2}}}{|2\kappa+1|}
\Omega_{\kappa+1,\mu}(\mathbf{n})
\nonumber \\
&& +i(\kappa+1)\frac{\sqrt{(\kappa-\frac{1}{2})^{2}-\mu^{2}}}
{|2\kappa-1|}\Omega_{\kappa-1,\mu}(\mathbf{n})
\label{3.2.11}
\end{eqnarray}
\begin{eqnarray}
\mathbf{e}_{\pm1}\cdot(\mathbf{n}\times\hat{\mathbf{L}})\,
\Omega_{\kappa\mu}(\mathbf{n})
&=& \pm i\sqrt{2}(\kappa+1)
\frac{\sqrt{\kappa^{2}-(\mu\pm\frac{1}{2})^{2}}}
{4\kappa^{2}-1}\Omega_{-\kappa,\mu\pm1}(\mathbf{n})
\nonumber \\
&& -i\kappa\frac{\sqrt{(\kappa\pm\mu+\frac{1}{2})
(\kappa\pm\mu+\frac{3}{2})}}{\sqrt{2}(2\kappa+1)}
\Omega_{\kappa+1,\mu\pm1}(\mathbf{n})
\nonumber \\
&& -i(\kappa+1)\frac{\sqrt{(\kappa\mp\mu-\frac{1}{2})
(\kappa\mp\mu-\frac{3}{2})}}{\sqrt{2}(2\kappa-1)}
\Omega_{\kappa-1,\mu\pm1}(\mathbf{n})
\label{3.2.12}
\end{eqnarray}
\begin{eqnarray}
\mathbf{e}_{0}\cdot(\hat{\mathbf{L}}\times\mathbf{n})\,
\Omega_{\kappa\mu}(\mathbf{n})
&=& i\frac{2\mu(\kappa-1)}{4\kappa^{2}-1}
\Omega_{-\kappa\mu}(\mathbf{n})
+i(\kappa+2)
\frac{\sqrt{(\kappa+\frac{1}{2})^{2}-\mu^{2}}}{|2\kappa+1|}
\Omega_{\kappa+1,\mu}(\mathbf{n})
\nonumber \\
&& -i(\kappa-1)\frac{\sqrt{(\kappa-\frac{1}{2})^{2}-\mu^{2}}}
{|2\kappa-1|}\Omega_{\kappa-1,\mu}(\mathbf{n})
\label{3.2.13}
\end{eqnarray}
\begin{eqnarray}
\mathbf{e}_{\pm1}\cdot(\hat{\mathbf{L}}\times\mathbf{n})\,
\Omega_{\kappa\mu}(\mathbf{n})
&=& \mp i\sqrt{2}(\kappa-1)
\frac{\sqrt{\kappa^{2}-(\mu\pm\frac{1}{2})^{2}}}
{4\kappa^{2}-1}\Omega_{-\kappa,\mu\pm1}(\mathbf{n})
\nonumber \\
&& +i(\kappa+2)\frac{\sqrt{(\kappa\pm\mu+\frac{1}{2})
(\kappa\pm\mu+\frac{3}{2})}}{\sqrt{2}(2\kappa+1)}
\Omega_{\kappa+1,\mu\pm1}(\mathbf{n})
\nonumber \\
&& +i(\kappa-1)\frac{\sqrt{(\kappa\mp\mu-\frac{1}{2})
(\kappa\mp\mu-\frac{3}{2})}}{\sqrt{2}(2\kappa-1)}
\Omega_{\kappa-1,\mu\pm1}(\mathbf{n})
\label{3.2.14}
\end{eqnarray}
\begin{equation}
\mathbf{n}\cdot(\boldsymbol{\sigma}\times\hat{\mathbf{L}})\,
\Omega_{\kappa\mu}(\mathbf{n})
=i(\kappa+1)\Omega_{-\kappa\mu}(\mathbf{n})
\label{3.2.15}
\end{equation}
\begin{equation}
\mathbf{e}_{0}\cdot(\boldsymbol{\sigma}\times\hat{\mathbf{L}})\,
\Omega_{\kappa\mu}(\mathbf{n})
=i\,\sgn(\kappa)\sqrt{(\kappa+{\textstyle\frac{1}{2}})^{2}-\mu^{2}}\,
\Omega_{-\kappa-1,\mu}(\mathbf{n})
\label{3.2.16}
\end{equation}
\begin{equation}
\mathbf{e}_{\pm1}\cdot(\boldsymbol{\sigma}\times\hat{\mathbf{L}})\,
\Omega_{\kappa\mu}(\mathbf{n})
=i\frac{1}{\sqrt{2}}
\sqrt{(\kappa\pm\mu+{\textstyle\frac{1}{2}})
(\kappa\pm\mu+{\textstyle\frac{3}{2}})}\,
\Omega_{-\kappa-1,\mu\pm1}(\mathbf{n})
\label{3.2.17}
\end{equation}
\begin{equation}
(\boldsymbol{\sigma}\times\hat{\mathbf{L}})\cdot\mathbf{n}\,
\Omega_{\kappa\mu}(\mathbf{n})=i(\kappa-1)
\Omega_{-\kappa\mu}(\mathbf{n})
\label{3.2.18}
\end{equation}
\begin{eqnarray}
\mathbf{e}_{0}\cdot(\mathbf{n}\times\hat{\mathbf{J}})\,
\Omega_{\kappa\mu}(\mathbf{n})
&=& -i\frac{2\mu}{4\kappa^{2}-1}\Omega_{-\kappa\mu}(\mathbf{n})
-i\left(\kappa-{\textstyle\frac{1}{2}}\right)
\frac{\sqrt{(\kappa+\frac{1}{2})^{2}-\mu^{2}}}{|2\kappa+1|}
\Omega_{\kappa+1,\mu}(\mathbf{n})
\nonumber \\
&& +i\left(\kappa+{\textstyle\frac{1}{2}}\right)
\frac{\sqrt{(\kappa-\frac{1}{2})^{2}-\mu^{2}}}{|2\kappa-1|}
\Omega_{\kappa-1,\mu}(\mathbf{n})
\label{3.2.19}
\end{eqnarray}
\begin{eqnarray}
\mathbf{e}_{\pm1}\cdot(\mathbf{n}\times\hat{\mathbf{J}})\,
\Omega_{\kappa\mu}(\mathbf{n})
&=& \pm i\sqrt{2}\frac{\sqrt{\kappa^{2}-(\mu\pm\frac{1}{2})^{2}}}
{4\kappa^{2}-1}\Omega_{-\kappa,\mu\pm1}(\mathbf{n})
\nonumber \\
&& -i\left(\kappa-{\textstyle\frac{1}{2}}\right)
\frac{\sqrt{(\kappa\pm\mu+\frac{1}{2})
(\kappa\pm\mu+\frac{3}{2})}}{\sqrt{2}(2\kappa+1)}
\Omega_{\kappa+1,\mu\pm1}(\mathbf{n})
\nonumber \\
&& -i\left(\kappa+{\textstyle\frac{1}{2}}\right)
\frac{\sqrt{(\kappa\mp\mu-\frac{1}{2})
(\kappa\mp\mu-\frac{3}{2})}}{\sqrt{2}(2\kappa-1)}
\Omega_{\kappa-1,\mu\pm1}(\mathbf{n})
\label{3.2.20}
\end{eqnarray}
\begin{eqnarray}
\mathbf{e}_{0}\cdot(\hat{\mathbf{J}}\times\mathbf{n})\,
\Omega_{\kappa\mu}(\mathbf{n})
&=& -i\frac{2\mu}{4\kappa^{2}-1}\Omega_{-\kappa\mu}(\mathbf{n})
+i\left(\kappa+{\textstyle\frac{3}{2}}\right)
\frac{\sqrt{(\kappa+\frac{1}{2})^{2}-\mu^{2}}}{|2\kappa+1|}
\Omega_{\kappa+1,\mu}(\mathbf{n})
\nonumber \\
&& -i\left(\kappa-{\textstyle\frac{3}{2}}\right)
\frac{\sqrt{(\kappa-\frac{1}{2})^{2}-\mu^{2}}}{|2\kappa-1|}
\Omega_{\kappa-1,\mu}(\mathbf{n})
\label{3.2.21}
\end{eqnarray}
\begin{eqnarray}
\mathbf{e}_{\pm1}\cdot(\hat{\mathbf{J}}\times\mathbf{n})\,
\Omega_{\kappa\mu}(\mathbf{n})
&=& \pm i\sqrt{2}\frac{\sqrt{\kappa^{2}-(\mu\pm\frac{1}{2})^{2}}}
{4\kappa^{2}-1}\Omega_{-\kappa,\mu\pm1}(\mathbf{n})
\nonumber \\
&& +i\left(\kappa+{\textstyle\frac{3}{2}}\right)
\frac{\sqrt{(\kappa\pm\mu+\frac{1}{2})(\kappa\pm\mu+\frac{3}{2})}}
{\sqrt{2}(2\kappa+1)}\Omega_{\kappa+1,\mu\pm1}(\mathbf{n})
\nonumber \\
&& +i\left(\kappa-{\textstyle\frac{3}{2}}\right)
\frac{\sqrt{(\kappa\mp\mu-\frac{1}{2})(\kappa\mp\mu-\frac{3}{2})}}
{\sqrt{2}(2\kappa-1)}\Omega_{\kappa-1,\mu\pm1}(\mathbf{n})
\label{3.2.22}
\end{eqnarray}
\begin{equation}
\mathbf{n}\cdot(\boldsymbol{\sigma}\times\hat{\mathbf{J}})\,
\Omega_{\kappa\mu}(\mathbf{n})=i\kappa\Omega_{-\kappa\mu}(\mathbf{n})
\label{3.2.23}
\end{equation}
\begin{equation}
\mathbf{e}_{0}\cdot(\boldsymbol{\sigma}\times\hat{\mathbf{J}})\,
\Omega_{\kappa\mu}(\mathbf{n})=-i\frac{2\mu}{2\kappa+1}
\Omega_{\kappa\mu}(\mathbf{n})
+i(2\kappa-1)\frac{\sqrt{(\kappa+\frac{1}{2})^{2}-\mu^{2}}}
{|2\kappa+1|}\Omega_{-\kappa-1,\mu}(\mathbf{n})
\label{3.2.24}
\end{equation}
\begin{eqnarray}
\mathbf{e}_{\pm1}\cdot(\boldsymbol{\sigma}\times\hat{\mathbf{J}})\,
\Omega_{\kappa\mu}(\mathbf{n})
&=& \pm i\sqrt{2}
\frac{\sqrt{\kappa^{2}-(\mu\pm\frac{1}{2})^{2}}}{2\kappa+1}
\Omega_{\kappa,\mu\pm1}(\mathbf{n})
\nonumber \\
&& +i(2\kappa-1)\frac{\sqrt{(\kappa\pm\mu+\frac{1}{2})
(\kappa\pm\mu+\frac{3}{2})}}{\sqrt{2}(2\kappa+1)}
\Omega_{-\kappa-1,\mu\pm1}(\mathbf{n})
\label{3.2.25}
\end{eqnarray}
\begin{equation}
(\boldsymbol{\sigma}\times\hat{\mathbf{J}})\cdot\mathbf{n}\,
\Omega_{\kappa\mu}(\mathbf{n})
=i(\kappa-2)\Omega_{-\kappa\mu}(\mathbf{n})
\label{3.2.26}
\end{equation}
\begin{equation}
\mathbf{n}\cdot(\hat{\mathbf{J}}\times\boldsymbol{\sigma})\,
\Omega_{\kappa\mu}(\mathbf{n})
=-i(\kappa+2)\Omega_{-\kappa\mu}(\mathbf{n})
\label{3.2.27}
\end{equation}
\begin{equation}
\mathbf{e}_{0}\cdot(\hat{\mathbf{J}}\times\boldsymbol{\sigma})\,
\Omega_{\kappa\mu}(\mathbf{n})
=-i\frac{2\mu}{2\kappa+1}\Omega_{\kappa\mu}(\mathbf{n})
-i(2\kappa+3)\frac{\sqrt{(\kappa+\frac{1}{2})^{2}-\mu^{2}}}
{|2\kappa+1|}\Omega_{-\kappa-1,\mu}(\mathbf{n})
\label{3.2.28}
\end{equation}
\begin{eqnarray}
\mathbf{e}_{\pm1}\cdot(\hat{\mathbf{J}}\times\boldsymbol{\sigma})\,
\Omega_{\kappa\mu}(\mathbf{n})
&=& \pm i\sqrt{2}\frac{\sqrt{\kappa^{2}-(\mu\pm\frac{1}{2})^{2}}}
{2\kappa+1}\Omega_{\kappa,\mu\pm1}(\mathbf{n})
\nonumber \\
&& -i(2\kappa+3)\frac{\sqrt{(\kappa\pm\mu+\frac{1}{2})
(\kappa\pm\mu+\frac{3}{2})}}{\sqrt{2}(2\kappa+1)}
\Omega_{-\kappa-1,\mu\pm1}(\mathbf{n})
\label{3.2.29}
\end{eqnarray}
\begin{equation}
(\hat{\mathbf{J}}\times\boldsymbol{\sigma})\cdot\mathbf{n}\,
\Omega_{\kappa\mu}(\mathbf{n})
=-i\kappa\Omega_{-\kappa\mu}(\mathbf{n})
\label{3.2.30}
\end{equation}
\begin{equation}
\mathbf{n}\cdot(\hat{\mathbf{L}}\times\hat{\mathbf{J}})\,
\Omega_{\kappa\mu}(\mathbf{n})=-i\frac{1}{2}(\kappa+1)
\Omega_{-\kappa\mu}(\mathbf{n})
\label{3.2.31}
\end{equation}
\begin{equation}
\mathbf{e}_{0}\cdot(\hat{\mathbf{L}}\times\hat{\mathbf{J}})\,
\Omega_{\kappa\mu}(\mathbf{n}) 
=i\frac{2\mu(\kappa+1)}{2\kappa+1}
\Omega_{\kappa\mu}(\mathbf{n})
-i\left(\kappa-{\textstyle\frac{1}{2}}\right)
\frac{\sqrt{(\kappa+\frac{1}{2})^{2}-\mu^{2}}}{|2\kappa+1|}
\Omega_{-\kappa-1,\mu}(\mathbf{n})
\label{3.2.32}
\end{equation}
\begin{eqnarray}
\mathbf{e}_{\pm1}\cdot(\hat{\mathbf{L}}\times\hat{\mathbf{J}})\,
\Omega_{\kappa\mu}(\mathbf{n})
&=& \mp i\sqrt{2}(\kappa+1)
\frac{\sqrt{\kappa^{2}-(\mu\pm\frac{1}{2})^{2}}}{2\kappa+1}
\Omega_{\kappa,\mu\pm1}(\mathbf{n})
\nonumber \\
&& -i\left(\kappa-{\textstyle\frac{1}{2}}\right)
\frac{\sqrt{(\kappa\pm\mu+\frac{1}{2})
(\kappa\pm\mu+\frac{3}{2})}}{\sqrt{2}(2\kappa+1)}
\Omega_{-\kappa-1,\mu\pm1}(\mathbf{n})
\label{3.2.33}
\end{eqnarray}
\begin{equation}
(\hat{\mathbf{L}}\times\hat{\mathbf{J}})\cdot\mathbf{n}\,
\Omega_{\kappa\mu}(\mathbf{n})
=-i\frac{1}{2}(\kappa-1)\Omega_{-\kappa\mu}(\mathbf{n})
\label{3.2.34}
\end{equation}
\begin{equation}
(\hat{\mathbf{L}}\times\hat{\mathbf{J}})\cdot\boldsymbol{\sigma}\,
\Omega_{\kappa\mu}(\mathbf{n})
=-i2(\kappa+1)\Omega_{\kappa\mu}(\mathbf{n})
\label{3.2.35}
\end{equation}
\begin{equation}
\mathbf{n}\cdot(\hat{\mathbf{J}}\times\hat{\mathbf{L}})\,
\Omega_{\kappa\mu}(\mathbf{n})
=i\frac{1}{2}(\kappa+1)\Omega_{-\kappa\mu}(\mathbf{n})
\label{3.2.36}
\end{equation}
\begin{equation}
\mathbf{e}_{0}\cdot(\hat{\mathbf{J}}\times\hat{\mathbf{L}})\,
\Omega_{\kappa\mu}(\mathbf{n})
=i\frac{2\mu(\kappa+1)}{2\kappa+1}\Omega_{\kappa\mu}(\mathbf{n})
+i\left(\kappa+{\textstyle\frac{3}{2}}\right)
\frac{\sqrt{(\kappa+\frac{1}{2})^{2}-\mu^{2}}}{|2\kappa+1|}
\Omega_{-\kappa-1,\mu}(\mathbf{n})
\label{3.2.37}
\end{equation}
\begin{eqnarray}
\mathbf{e}_{\pm1}\cdot(\hat{\mathbf{J}}\times\hat{\mathbf{L}})\,
\Omega_{\kappa\mu}(\mathbf{n})
&=& \mp i\sqrt{2}(\kappa+1)
\frac{\sqrt{\kappa^{2}-(\mu\pm\frac{1}{2})^{2}}}{2\kappa+1}
\Omega_{\kappa,\mu\pm1}(\mathbf{n})
\nonumber \\
&& +i\left(\kappa+{\textstyle\frac{3}{2}}\right)
\frac{\sqrt{(\kappa\pm\mu+\frac{1}{2})(\kappa\pm\mu+\frac{3}{2})}}
{\sqrt{2}(2\kappa+1)}\Omega_{-\kappa-1,\mu\pm1}(\mathbf{n})
\label{3.2.38}
\end{eqnarray}
\begin{equation}
\boldsymbol{\sigma}\cdot(\hat{\mathbf{J}}\times\hat{\mathbf{L}})\,
\Omega_{\kappa\mu}(\mathbf{n})
=-i2(\kappa+1)\Omega_{\kappa\mu}(\mathbf{n})
\label{3.2.39}
\end{equation}
\begin{equation}
(\hat{\mathbf{J}}\times\hat{\mathbf{L}})\cdot\mathbf{n}\,
\Omega_{\kappa\mu}(\mathbf{n})
=i\frac{1}{2}(\kappa-1)\Omega_{-\kappa\mu}(\mathbf{n})
\label{3.2.40}
\end{equation}
\begin{equation}
\hat{\mathbf{L}}\cdot(\mathbf{n}\times\hat{\mathbf{J}})\,
\Omega_{\kappa\mu}(\mathbf{n})
=i\frac{1}{2}(\kappa-1)\Omega_{-\kappa\mu}(\mathbf{n})
\label{3.2.41}
\end{equation}
\begin{equation}
\hat{\mathbf{J}}\cdot(\mathbf{n}\times\hat{\mathbf{L}})\,
\Omega_{\kappa\mu}(\mathbf{n})
=-i\frac{1}{2}(\kappa+1)\Omega_{-\kappa\mu}(\mathbf{n})
\label{3.2.42}
\end{equation}
\begin{equation}
\boldsymbol{\sigma}\cdot(\hat{\mathbf{J}}\times\boldsymbol{\sigma})\,
\Omega_{\kappa\mu}(\mathbf{n})
=i(2\kappa+5)\Omega_{\kappa\mu}(\mathbf{n})
\label{3.2.43}
\end{equation}
\begin{equation}
\hat{\mathbf{J}}\cdot(\mathbf{n}\times\hat{\mathbf{J}})\,
\Omega_{\kappa\mu}(\mathbf{n})
=-i\frac{1}{2}\Omega_{-\kappa\mu}(\mathbf{n})
\label{3.2.44}
\end{equation}
\begin{equation}
\hat{\mathbf{J}}\cdot(\boldsymbol{\sigma}\times\hat{\mathbf{J}})\,
\Omega_{\kappa\mu}(\mathbf{n})
=-i\left(\kappa-{\textstyle\frac{1}{2}}\right)
\Omega_{\kappa\mu}(\mathbf{n})
\label{3.2.45}
\end{equation}
\begin{equation}
\hat{\mathbf{J}}\cdot(\hat{\mathbf{L}}\times\hat{\mathbf{J}})\,
\Omega_{\kappa\mu}(\mathbf{n})
=i\left(\kappa-{\textstyle\frac{1}{2}}\right)
(\kappa+1)\Omega_{\kappa\mu}(\mathbf{n})
\label{3.2.46}
\end{equation}
\begin{equation}
(\hat{\mathbf{L}}\times\hat{\mathbf{J}})\cdot\hat{\mathbf{L}}\,
\Omega_{\kappa\mu}(\mathbf{n})
=i\left(\kappa+{\textstyle\frac{1}{2}}\right)
(\kappa+1)\Omega_{\kappa\mu}(\mathbf{n})
\label{3.2.47}
\end{equation}
%
%
\subsection{Differential relations of the second kind}
\label{III.3}
\setcounter{equation}{0}
\begin{equation}
\mathbf{n}\cdot\boldsymbol{\nabla}\,
F(r)\Omega_{\kappa\mu}(\mathbf{n})
=\frac{\mathrm{d}F(r)}{\mathrm{d}r}\Omega_{\kappa\mu}(\mathbf{n})
\label{3.3.1}
\end{equation}
\begin{equation}
\boldsymbol{\nabla}\cdot\mathbf{n}\,
F(r)\Omega_{\kappa\mu}(\mathbf{n})
=\left(\frac{\partial}{\partial r}+\frac{2}{r}\right)F(r)
\Omega_{\kappa\mu}(\mathbf{n})
\label{3.3.2}
\end{equation}
\begin{eqnarray}
\mathbf{e}_{0}\cdot\boldsymbol{\nabla}\,
F(r)\Omega_{\kappa\mu}(\mathbf{n})
&=& -\frac{2\mu}{4\kappa^{2}-1}
\left(\frac{\partial}{\partial r}+\frac{\kappa+1}{r}\right)F(r)
\Omega_{-\kappa\mu}(\mathbf{n})
\nonumber \\
&& +\frac{\sqrt{(\kappa+\frac{1}{2})^{2}-\mu^{2}}}{|2\kappa+1|}
\left(\frac{\partial}{\partial r}-\frac{\kappa}{r}\right)F(r)
\Omega_{\kappa+1,\mu}(\mathbf{n})
\nonumber \\
&& +\frac{\sqrt{(\kappa-\frac{1}{2})^{2}-\mu^{2}}}{|2\kappa-1|}
\left(\frac{\partial}{\partial r}+\frac{\kappa+1}{r}\right)F(r)
\Omega_{\kappa-1,\mu}(\mathbf{n})
\label{3.3.3}
\end{eqnarray}
\begin{eqnarray}
\mathbf{e}_{\pm1}\cdot\boldsymbol{\nabla}\,
F(r)\Omega_{\kappa\mu}(\mathbf{n})
&=& \pm\sqrt{2}\frac{\sqrt{\kappa^{2}-(\mu\pm\frac{1}{2})^{2}}}
{4\kappa^{2}-1}\left(\frac{\partial}{\partial r}
+\frac{\kappa+1}{r}\right)F(r)\Omega_{-\kappa,\mu\pm1}(\mathbf{n})
\nonumber \\
&& +\frac{\sqrt{(\kappa\pm\mu+\frac{1}{2})(\kappa\pm\mu+\frac{3}{2})}}
{\sqrt{2}(2\kappa+1)}
\left(\frac{\partial}{\partial r}-\frac{\kappa}{r}\right)F(r)
\Omega_{\kappa+1,\mu\pm1}(\mathbf{n})
\nonumber \\
&&
-\frac{\sqrt{(\kappa\mp\mu-\frac{1}{2})(\kappa\mp\mu-\frac{3}{2})}}
{\sqrt{2}(2\kappa-1)}
\left(\frac{\partial}{\partial r}+\frac{\kappa+1}{r}\right)F(r)
\Omega_{\kappa-1,\mu\pm1}(\mathbf{n})
\nonumber \\
&&
\label{3.3.4}
\end{eqnarray}
\begin{equation}
\boldsymbol{\sigma}\cdot\boldsymbol{\nabla}\,
F(r)\Omega_{\kappa\mu}(\mathbf{n})
=-\left(\frac{\partial}{\partial r}+\frac{\kappa+1}{r}\right)F(r)
\Omega_{-\kappa\mu}(\mathbf{n})
\label{3.3.5}
\end{equation}
\begin{equation}
\hat{\mathbf{J}}\cdot\boldsymbol{\nabla}\,
F(r)\Omega_{\kappa\mu}(\mathbf{n})
=-\frac{1}{2}\left(\frac{\partial}{\partial r}
+\frac{\kappa+1}{r}\right)F(r)\Omega_{-\kappa\mu}(\mathbf{n})
\label{3.3.6}
\end{equation}
\begin{equation}
\boldsymbol{\nabla}\cdot\hat{\mathbf{J}}\,
F(r)\Omega_{\kappa\mu}(\mathbf{n})
=-\frac{1}{2}\left(\frac{\partial}{\partial r}
+\frac{\kappa+1}{r}\right)F(r)\Omega_{-\kappa\mu}(\mathbf{n})
\label{3.3.7}
\end{equation}
\begin{equation}
\boldsymbol{\nabla}^{2}\,
F(r)\Omega_{\kappa\mu}(\mathbf{n})
=\frac{1}{r}\left(\frac{\partial^{2}}{\partial r^{2}}
-\frac{\kappa(\kappa+1)}{r^{2}}\right)
rF(r)\Omega_{\kappa\mu}(\mathbf{n})
\label{3.3.8}
\end{equation}
\begin{eqnarray}
\mathbf{e}_{0}
\cdot(\boldsymbol{\sigma}\times\boldsymbol{\nabla})\,
F(r)\Omega_{\kappa\mu}(\mathbf{n})
&=& -i\frac{4\mu\kappa}{4\kappa^{2}-1}
\left(\frac{\partial}{\partial r}
+\frac{\kappa+1}{r}\right)F(r)\Omega_{-\kappa\mu}(\mathbf{n})
\nonumber \\
&& -i\frac{\sqrt{(\kappa+\frac{1}{2})^{2}-\mu^{2}}}
{|2\kappa+1|}\left(\frac{\partial}{\partial r}
-\frac{\kappa}{r}\right)F(r)\Omega_{\kappa+1,\mu}(\mathbf{n})
\nonumber \\
&& +i\frac{\sqrt{(\kappa-\frac{1}{2})^{2}-\mu^{2}}}
{|2\kappa-1|}\left(\frac{\partial}{\partial r}
+\frac{\kappa+1}{r}\right)F(r)\Omega_{\kappa-1,\mu}(\mathbf{n})
\label{3.3.9}
\end{eqnarray}
\begin{eqnarray}
\mathbf{e}_{\pm1}\cdot(\boldsymbol{\sigma}\times\boldsymbol{\nabla})\,
F(r)\Omega_{\kappa\mu}(\mathbf{n})
&=& \pm i2\sqrt{2}\kappa
\frac{\sqrt{\kappa^{2}-(\mu\pm\frac{1}{2})^{2}}}{4\kappa^{2}-1}
\left(\frac{\partial}{\partial r}+\frac{\kappa+1}{r}\right)
F(r)\Omega_{-\kappa,\mu\pm1}(\mathbf{n})
\nonumber \\
&& -i\frac{\sqrt{(\kappa\pm\mu+\frac{1}{2})
(\kappa\pm\mu+\frac{3}{2})}}{\sqrt{2}(2\kappa+1)}
\left(\frac{\partial}{\partial r}-\frac{\kappa}{r}\right)
F(r)\Omega_{\kappa+1,\mu\pm1}(\mathbf{n})
\nonumber \\
&& -i\frac{\sqrt{(\kappa\mp\mu-\frac{1}{2})
(\kappa\mp\mu-\frac{3}{2})}}{\sqrt{2}(2\kappa-1)}
\left(\frac{\partial}{\partial r}+\frac{\kappa+1}{r}\right)
F(r)\Omega_{\kappa-1,\mu\pm1}(\mathbf{n})
\nonumber \\
&&
\label{3.3.10}
\end{eqnarray}
\begin{equation}
\hat{\mathbf{J}}\cdot(\boldsymbol{\sigma}\times\boldsymbol{\nabla})\,
F(r)\Omega_{\kappa\mu}(\mathbf{n})
=-i\kappa\left(\frac{\partial}{\partial r}+\frac{\kappa+1}{r}\right)
F(r)\Omega_{-\kappa\mu}(\mathbf{n})
\label{3.3.11}
\end{equation}
\begin{equation}
\mathbf{n}\cdot(\hat{\mathbf{L}}\times\boldsymbol{\nabla})\,
F(r)\Omega_{\kappa\mu}(\mathbf{n})
=i\left(2\frac{\partial}{\partial r}-\frac{\kappa(\kappa+1)}{r}\right)
F(r)\Omega_{\kappa\mu}(\mathbf{n})
\label{3.3.12}
\end{equation}
\begin{eqnarray}
\mathbf{e}_{0}\cdot(\hat{\mathbf{L}}\times\boldsymbol{\nabla})\,
F(r)\Omega_{\kappa\mu}(\mathbf{n})
&=& i\frac{2\mu(\kappa-1)}{4\kappa^{2}-1}
\left(\frac{\partial}{\partial r}+\frac{\kappa+1}{r}\right)
F(r)\Omega_{-\kappa\mu}(\mathbf{n})
\nonumber \\
&& +i(\kappa+2)
\frac{\sqrt{(\kappa+\frac{1}{2})^{2}-\mu^{2}}}{|2\kappa+1|}
\left(\frac{\partial}{\partial r}-\frac{\kappa}{r}\right)
F(r)\Omega_{\kappa+1,\mu}(\mathbf{n})
\nonumber \\
&& -i(\kappa-1)\frac{\sqrt{(\kappa-\frac{1}{2})^{2}-\mu^{2}}}
{|2\kappa-1|}
\left(\frac{\partial}{\partial r}+\frac{\kappa+1}{r}\right)
F(r)\Omega_{\kappa-1,\mu}(\mathbf{n})
\nonumber \\
&&
\label{3.3.13}
\end{eqnarray}
\begin{eqnarray}
\mathbf{e}_{\pm1}\cdot(\hat{\mathbf{L}}\times\boldsymbol{\nabla})\,
F(r)\Omega_{\kappa\mu}(\mathbf{n})
&=& \mp i\sqrt{2}(\kappa-1)
\frac{\sqrt{\kappa^{2}-(\mu\pm\frac{1}{2})^{2}}}
{4\kappa^{2}-1}
\left(\frac{\partial}{\partial r}+\frac{\kappa+1}{r}\right)
F(r)\Omega_{-\kappa,\mu\pm1}(\mathbf{n})
\nonumber \\
&& +i(\kappa+2)\frac{\sqrt{(\kappa\pm\mu+\frac{1}{2})
(\kappa\pm\mu+\frac{3}{2})}}{\sqrt{2}(2\kappa+1)}
\left(\frac{\partial}{\partial r}-\frac{\kappa}{r}\right)
F(r)\Omega_{\kappa+1,\mu\pm1}(\mathbf{n})
\nonumber \\
&& +i(\kappa-1)\frac{\sqrt{(\kappa\mp\mu-\frac{1}{2})
(\kappa\mp\mu-\frac{3}{2})}}{\sqrt{2}(2\kappa-1)}
\left(\frac{\partial}{\partial r}+\frac{\kappa+1}{r}\right)
F(r)\Omega_{\kappa-1,\mu\pm1}(\mathbf{n})
\nonumber \\
&&
\label{3.3.14}
\end{eqnarray}
\begin{equation}
\boldsymbol{\sigma}\cdot(\hat{\mathbf{L}}\times\boldsymbol{\nabla})\,
F(r)\Omega_{\kappa\mu}(\mathbf{n})
=i(\kappa-1)\left(\frac{\partial}{\partial r}
+\frac{\kappa+1}{r}\right)F(r)\Omega_{-\kappa\mu}(\mathbf{n})
\label{3.3.15}
\end{equation}
\begin{equation}
\hat{\mathbf{J}}\cdot(\hat{\mathbf{L}}\times\boldsymbol{\nabla})\,
F(r)\Omega_{\kappa\mu}(\mathbf{n})
=i\frac{1}{2}(\kappa-1)\left(\frac{\partial}{\partial r}
+\frac{\kappa+1}{r}\right)F(r)\Omega_{-\kappa\mu}(\mathbf{n})
\label{3.3.16}
\end{equation}
\begin{eqnarray}
\mathbf{e}_{0}\cdot(\boldsymbol{\nabla}\times\hat{\mathbf{L}})\,
F(r)\Omega_{\kappa\mu}(\mathbf{n})
&=& -i\frac{2\mu(\kappa+1)}{4\kappa^{2}-1}
\left(\frac{\partial}{\partial r}+\frac{\kappa+1}{r}\right)
F(r)\Omega_{-\kappa\mu}(\mathbf{n})
\nonumber \\
&& -i\kappa\frac{\sqrt{(\kappa+\frac{1}{2})^{2}-\mu^{2}}}{|2\kappa+1|}
\left(\frac{\partial}{\partial r}-\frac{\kappa}{r}\right)
F(r)\Omega_{\kappa+1,\mu}(\mathbf{n})
\nonumber \\
&& +i(\kappa+1)\frac{\sqrt{(\kappa-\frac{1}{2})^{2}-\mu^{2}}}
{|2\kappa-1|}
\left(\frac{\partial}{\partial r}+\frac{\kappa+1}{r}\right)
F(r)\Omega_{\kappa-1,\mu}(\mathbf{n})
\nonumber \\
&&
\label{3.3.17}
\end{eqnarray}
\begin{eqnarray}
\mathbf{e}_{\pm1}\cdot(\boldsymbol{\nabla}\times\hat{\mathbf{L}})\,
F(r)\Omega_{\kappa\mu}(\mathbf{n})
&=& \pm i\sqrt{2}(\kappa+1)
\frac{\sqrt{\kappa^{2}-(\mu\pm\frac{1}{2})^{2}}}
{4\kappa^{2}-1}
\left(\frac{\partial}{\partial r}+\frac{\kappa+1}{r}\right)
F(r)\Omega_{-\kappa,\mu\pm1}(\mathbf{n})
\nonumber \\
&& -i\kappa\frac{\sqrt{(\kappa\pm\mu+\frac{1}{2})
(\kappa\pm\mu+\frac{3}{2})}}{\sqrt{2}(2\kappa+1)}
\left(\frac{\partial}{\partial r}-\frac{\kappa}{r}\right)
F(r)\Omega_{\kappa+1,\mu\pm1}(\mathbf{n})
\nonumber \\
&& -i(\kappa+1)\frac{\sqrt{(\kappa\mp\mu-\frac{1}{2})
(\kappa\mp\mu-\frac{3}{2})}}{\sqrt{2}(2\kappa-1)}
\left(\frac{\partial}{\partial r}+\frac{\kappa+1}{r}\right)
F(r)\Omega_{\kappa-1,\mu\pm1}(\mathbf{n})
\nonumber \\
&&
\label{3.3.18}
\end{eqnarray}
\begin{equation}
\boldsymbol{\sigma}\cdot(\boldsymbol{\nabla}\times\hat{\mathbf{L}})\,
F(r)\Omega_{\kappa\mu}(\mathbf{n})
=-i(\kappa+1)\left(\frac{\partial}{\partial r}
+\frac{\kappa+1}{r}\right)F(r)\Omega_{-\kappa\mu}(\mathbf{n})
\label{3.3.19}
\end{equation}
\begin{equation}
\hat{\mathbf{J}}\cdot(\boldsymbol{\nabla}\times\hat{\mathbf{L}})\,
F(r)\Omega_{\kappa\mu}(\mathbf{n})
=-i\frac{1}{2}(\kappa+1)\left(\frac{\partial}{\partial r}
+\frac{\kappa+1}{r}\right)F(r)\Omega_{-\kappa\mu}(\mathbf{n})
\label{3.3.20}
\end{equation}
\begin{equation}
(\boldsymbol{\nabla}\times\hat{\mathbf{L}})\cdot\mathbf{n}\,
F(r)\Omega_{\kappa\mu}(\mathbf{n})
=i\left(2\frac{\partial}{\partial r}
+\frac{\kappa^{2}+\kappa+4}{r}\right)
F(r)\Omega_{\kappa\mu}(\mathbf{n})
\label{3.3.21}
\end{equation}
\begin{equation}
(\boldsymbol{\nabla}\times\hat{\mathbf{L}})\cdot\hat{\mathbf{J}}\,
F(r)\Omega_{\kappa\mu}(\mathbf{n})
=-i\frac{1}{2}(\kappa+1)\left(\frac{\partial}{\partial r}
+\frac{\kappa+1}{r}\right)
F(r)\Omega_{-\kappa\mu}(\mathbf{n})
\label{3.3.22}
\end{equation}
\begin{equation}
\mathbf{n}\cdot(\hat{\mathbf{J}}\times\boldsymbol{\nabla})\,
F(r)\Omega_{\kappa\mu}(\mathbf{n})
=i\left(2\frac{\partial}{\partial r}
-\frac{(\kappa+1)(2\kappa-1)}{2r}\right)
F(r)\Omega_{\kappa\mu}(\mathbf{n})
\label{3.3.23}
\end{equation}
\begin{eqnarray}
\mathbf{e}_{0}\cdot(\hat{\mathbf{J}}\times\boldsymbol{\nabla})\,
F(r)\Omega_{\kappa\mu}(\mathbf{n})
&=& -i\frac{2\mu}{4\kappa^{2}-1}
\left(\frac{\partial}{\partial r}+\frac{\kappa+1}{r}\right)
F(r)\Omega_{-\kappa\mu}(\mathbf{n})
\nonumber \\
&& +i\left(\kappa+{\textstyle\frac{3}{2}}\right)
\frac{\sqrt{(\kappa+\frac{1}{2})^{2}-\mu^{2}}}{|2\kappa+1|}
\left(\frac{\partial}{\partial r}-\frac{\kappa}{r}\right)
F(r)\Omega_{\kappa+1,\mu}(\mathbf{n})
\nonumber \\
&& -i\left(\kappa-{\textstyle\frac{3}{2}}\right)
\frac{\sqrt{(\kappa-\frac{1}{2})^{2}-\mu^{2}}}{|2\kappa-1|}
\left(\frac{\partial}{\partial r}+\frac{\kappa+1}{r}\right)
F(r)\Omega_{\kappa-1,\mu}(\mathbf{n})
\nonumber \\
&&
\label{3.3.24}
\end{eqnarray}
\begin{eqnarray}
\mathbf{e}_{\pm1}\cdot(\hat{\mathbf{J}}\times\boldsymbol{\nabla})\,
F(r)\Omega_{\kappa\mu}(\mathbf{n})
&=& \pm i\sqrt{2}\frac{\sqrt{\kappa^{2}-(\mu\pm\frac{1}{2})^{2}}}
{4\kappa^{2}-1}
\left(\frac{\partial}{\partial r}+\frac{\kappa+1}{r}\right)
F(r)\Omega_{-\kappa,\mu\pm1}(\mathbf{n})
\nonumber \\
&& +i\left(\kappa+{\textstyle\frac{3}{2}}\right)
\frac{\sqrt{(\kappa\pm\mu+\frac{1}{2})(\kappa\pm\mu+\frac{3}{2})}}
{\sqrt{2}(2\kappa+1)}
\left(\frac{\partial}{\partial r}-\frac{\kappa}{r}\right)
F(r)\Omega_{\kappa+1,\mu\pm1}(\mathbf{n})
\nonumber \\
&& +i\left(\kappa-{\textstyle\frac{3}{2}}\right)
\frac{\sqrt{(\kappa\mp\mu-\frac{1}{2})(\kappa\mp\mu-\frac{3}{2})}}
{\sqrt{2}(2\kappa-1)}
\left(\frac{\partial}{\partial r}+\frac{\kappa+1}{r}\right)
F(r)\Omega_{\kappa-1,\mu\pm1}(\mathbf{n})
\nonumber \\
&&
\label{3.3.25}
\end{eqnarray}
\begin{equation}
\boldsymbol{\sigma}\cdot(\hat{\mathbf{J}}\times\boldsymbol{\nabla})\,
F(r)\Omega_{\kappa\mu}(\mathbf{n})
=i(\kappa-2)\left(\frac{\partial}{\partial r}
+\frac{\kappa+1}{r}\right)F(r)\Omega_{-\kappa\mu}(\mathbf{n})
\label{3.3.26}
\end{equation}
\begin{equation}
\hat{\mathbf{L}}\cdot(\hat{\mathbf{J}}\times\boldsymbol{\nabla})\,
F(r)\Omega_{\kappa\mu}(\mathbf{n})
=-i\frac{1}{2}(\kappa-1)\left(\frac{\partial}{\partial r}
+\frac{\kappa+1}{r}\right)F(r)\Omega_{-\kappa\mu}(\mathbf{n})
\label{3.3.27}
\end{equation}
\begin{eqnarray}
\mathbf{e}_{0}\cdot(\boldsymbol{\nabla}\times\hat{\mathbf{J}})\,
F(r)\Omega_{\kappa\mu}(\mathbf{n})
&=& -i\frac{2\mu}{4\kappa^{2}-1}
\left(\frac{\partial}{\partial r}+\frac{\kappa+1}{r}\right)
F(r)\Omega_{-\kappa\mu}(\mathbf{n})
\nonumber \\
&& -i\left(\kappa-{\textstyle\frac{1}{2}}\right)
\frac{\sqrt{(\kappa+\frac{1}{2})^{2}-\mu^{2}}}{|2\kappa+1|}
\left(\frac{\partial}{\partial r}-\frac{\kappa}{r}\right)
F(r)\Omega_{\kappa+1,\mu}(\mathbf{n})
\nonumber \\
&& +i\left(\kappa+{\textstyle\frac{1}{2}}\right)
\frac{\sqrt{(\kappa-\frac{1}{2})^{2}-\mu^{2}}}{|2\kappa-1|}
\left(\frac{\partial}{\partial r}+\frac{\kappa+1}{r}\right)
F(r)\Omega_{\kappa-1,\mu}(\mathbf{n})
\nonumber \\
&&
\label{3.3.28}
\end{eqnarray}
\begin{eqnarray}
\mathbf{e}_{\pm1}\cdot(\boldsymbol{\nabla}\times\hat{\mathbf{J}})\,
F(r)\Omega_{\kappa\mu}(\mathbf{n})
&=& \pm i\sqrt{2}\frac{\sqrt{\kappa^{2}-(\mu\pm\frac{1}{2})^{2}}}
{4\kappa^{2}-1}
\left(\frac{\partial}{\partial r}+\frac{\kappa+1}{r}\right)
F(r)\Omega_{-\kappa,\mu\pm1}(\mathbf{n})
\nonumber \\
&& -i\left(\kappa-{\textstyle\frac{1}{2}}\right)
\frac{\sqrt{(\kappa\pm\mu+\frac{1}{2})
(\kappa\pm\mu+\frac{3}{2})}}{\sqrt{2}(2\kappa+1)}
\left(\frac{\partial}{\partial r}-\frac{\kappa}{r}\right)
F(r)\Omega_{\kappa+1,\mu\pm1}(\mathbf{n})
\nonumber \\
&& -i\left(\kappa+{\textstyle\frac{1}{2}}\right)
\frac{\sqrt{(\kappa\mp\mu-\frac{1}{2})
(\kappa\mp\mu-\frac{3}{2})}}{\sqrt{2}(2\kappa-1)}
\left(\frac{\partial}{\partial r}+\frac{\kappa+1}{r}\right)
F(r)\Omega_{\kappa-1,\mu\pm1}(\mathbf{n})
\nonumber \\
&&
\label{3.3.29}
\end{eqnarray}
\begin{equation}
\boldsymbol{\sigma}\cdot(\boldsymbol{\nabla}\times\hat{\mathbf{J}})\,
F(r)\Omega_{\kappa\mu}(\mathbf{n})
=-i\kappa\left(\frac{\partial}{\partial r}+\frac{\kappa+1}{r}\right)
F(r)\Omega_{-\kappa\mu}(\mathbf{n})
\label{3.3.30}
\end{equation}
\begin{equation}
\hat{\mathbf{L}}\cdot(\boldsymbol{\nabla}\times\hat{\mathbf{J}})\,
F(r)\Omega_{\kappa\mu}(\mathbf{n})=i\frac{1}{2}(\kappa-1)
\left(\frac{\partial}{\partial r}+\frac{\kappa+1}{r}\right)
F(r)\Omega_{-\kappa\mu}(\mathbf{n})
\label{3.3.31}
\end{equation}
\begin{equation}
\hat{\mathbf{J}}\cdot(\boldsymbol{\nabla}\times\hat{\mathbf{J}})\,
F(r)\Omega_{\kappa\mu}(\mathbf{n})=-i\frac{1}{2}
\left(\frac{\partial}{\partial r}+\frac{\kappa+1}{r}\right)
F(r)\Omega_{-\kappa\mu}(\mathbf{n})
\label{3.3.32}
\end{equation}
\begin{equation}
(\boldsymbol{\nabla}\times\hat{\mathbf{J}})\cdot\mathbf{n}\,
F(r)\Omega_{\kappa\mu}(\mathbf{n})
=i\left(2\frac{\partial}{\partial r}
+\frac{2\kappa^{2}+\kappa+7}{2r}\right)
F(r)\Omega_{\kappa\mu}(\mathbf{n})
\label{3.3.33}
\end{equation}
\begin{equation}
(\boldsymbol{\nabla}\times\hat{\mathbf{J}})\cdot\boldsymbol{\sigma}\,
F(r)\Omega_{\kappa\mu}(\mathbf{n})
=-i(\kappa+2)\left(\frac{\partial}{\partial r}
+\frac{\kappa+1}{r}\right)F(r)\Omega_{-\kappa\mu}(\mathbf{n})
\label{3.3.34}
\end{equation}
\begin{equation}
(\boldsymbol{\nabla}\times\hat{\mathbf{J}})\cdot\hat{\mathbf{L}}\,
F(r)\Omega_{\kappa\mu}(\mathbf{n})
=i\frac{1}{2}(\kappa+1)\left(\frac{\partial}{\partial r}
+\frac{\kappa+1}{r}\right)F(r)\Omega_{-\kappa\mu}(\mathbf{n})
\label{3.3.35}
\end{equation}
\begin{equation}
\boldsymbol{\nabla}\cdot(\hat{\mathbf{L}}\times\boldsymbol{\nabla})\,
F(r)\Omega_{\kappa\mu}(\mathbf{n})
=i\frac{2}{r}\left(\frac{\partial^{2}}{\partial r^{2}}
-\frac{\kappa(\kappa+1)}{r^{2}}\right)
rF(r)\Omega_{\kappa\mu}(\mathbf{n})
\label{3.3.36}
\end{equation}
\begin{equation}
\boldsymbol{\nabla}\cdot(\hat{\mathbf{J}}\times\boldsymbol{\nabla})\,
F(r)\Omega_{\kappa\mu}(\mathbf{n})
=i\frac{2}{r}\left(\frac{\partial^{2}}{\partial r^{2}}
-\frac{\kappa(\kappa+1)}{r^{2}}\right)
rF(r)\Omega_{\kappa\mu}(\mathbf{n})
\label{3.3.37}
\end{equation}
\section*{Acknowledgments}
The author acknowledges discussions with Sebastian Bielski, Justyna
Kunicka, and Krzysztof Mielewczyk.
%
%

%
\end{document}